\documentclass[fleqn,usenatbib,useAMS]{mnras}

\newcommand{\hei}{He\,I}
\newcommand{\lyalpha}{Ly-$\alpha$}

\newcommand{\mgi}{Mg\,I}
\newcommand{\sii}{Si\,I}

\newcommand{\tess}{\emph{TESS}}

\newcommand{\gaia}{\emph{Gaia}}
\newcommand{\ktwo}{\emph{K2}}

\newcommand{\kepler}{\emph{Kepler}}

\newcommand{\wise}{\emph{WISE}}

\newcommand{\kms}{\mbox{km\,s$^{-1}~$}}
\newcommand{\ks}{\mbox{km\,s$^{-1}~$}}

\newcommand{\msun}{M$_{\odot}$}

\newcommand{\rsun}{R$_{\odot}$}
\newcommand{\lsun}{L$_{\odot}~$}

\newcommand{\mearth}{$M_{\oplus}$}

\newcommand{\rearth}{$R_{\oplus}$}

\newcommand{\teff}{\ensuremath{T_{\rm eff}}}

\newcommand{\ebv}{E($B$-$V$)}

\newcommand{\hoststar}{V1298\,Tau}
\newcommand{\planet}{V1298\,Tau\,b}

\usepackage{newtxtext,newtxmath,lineno}

\usepackage[T1]{fontenc}
\usepackage{ae,aecompl}

\usepackage{graphicx}	
\usepackage{amsmath}	
\usepackage{amssymb}	

\title[]{Zodiacal Exoplanets in Time (ZEIT) XIII: Planet Orbits and Atmospheres in the \hoststar\ System, a Keystone in Studies of Early Planetary Evolution}

\author[Gaidos et al.]{
E. Gaidos\thanks{E-mail: gaidos@hawaii.edu}$^{1,2,3}$, T. Hirano$^{4,5,6}$, C. Beichman$^{7}$, J. Livingston$^{8}$,
H. Harakawa$^{9}$, 
\newauthor
K. W. Hodapp$^{10}$, 
M. Ishizuka$^{8}$,
S. Jacobson$^{10}$,
M. Konishi$^{11}$, 
T. Kotani$^{4,5,6}$,
T. Kudo$^{9}$, 
\newauthor
T. Kurokawa$^{4,12}$,
M. Kuzuhara$^{4,5}$, 
J. Nishikawa$^{5,6,4}$,
M. Omiya$^{4,5}$,
T. Serizawa$^{5,12}$,
\newauthor
M. Tamura$^{4,5,8}$, 
A. Ueda$^{4,5,6}$,
S. Vievard$^{9}$
\\
$^{1}$Department of Earth Sciences, University of Hawai'i at M\={a}noa, Honolulu, HI  96822, USA\\
$^{2}$Center for Space and Habitability, University of Bern, Gesellschaftsstrasse 6, 3012 Bern, Switzerland\\
$^{3}$Institute for Astrophysics, University of Vienna, T\"{u}rkenschanzstrasse 17, 1180 Vienna, Austria\\
$^{4}$Astrobiology Center, 2-21-1 Osawa, Mitaka, Tokyo 181-8588, Japan\\
$^{5}$National Astronomical Observatory of Japan, NINS, 2-21-1 Osawa, Mitaka, Tokyo 181-8588, Japan\\
$^{6}$Department of Astronomical Science, School of Physical Sciences, The Graduate University for Advanced Studies (SOKENDAI), 2-21-1, Osawa, Mitaka, Tokyo, 181-8588, Japan\\
$^{7}$IPAC, Caltech, Pasadena, CA 91125 USA\\
$^{8}$Department of Astronomy, Graduate School of Science, The University of Tokyo, 7-3-1 Hongo, Bunkyo-ku, Tokyo 113-0033, Japan\\
$^{9}$Subaru Telescope, 650 N. Aohoku Place, Hilo, HI 96720, USA\\
$^{10}$University of Hawaii, Institute for Astronomy, 640 N. Aohoku Place, Hilo, HI 96720, USA\\
$^{11}$Faculty of Science and Technology, Oita University, 700 Dannoharu, Oita 870-1192, Japan\\
$^{12}$Institute of Engineering, Tokyo University of Agriculture and Technology, 2-24-16, Nakacho, Koganei, Tokyo, 184-8588, Japan
}

\date{MNRAS.  Accepted 2021 October 20. Received 2021 October 19; in original form 2021 August 30}

\pubyear{2021}

\begin{document}
\label{firstpage}
\pagerange{\pageref{firstpage}--\pageref{lastpage}}
\maketitle

\begin{abstract}
Studies of planetary systems of stars in star-forming regions and young clusters open a window on the formative stages of planetary evolution.  We obtained high-cadence high-resolution infrared spectroscopy of the solar-mass Taurus association-member \hoststar\ during a transit of its 10\rearth-size ``b" planet.  We measured the systemic radial velocity and find that the kinematics of \hoststar\ suggest an affiliation with a $\gtrsim$6\,Myr-old subgroup.  A comparison of \hoststar\ and the nearby, co-moving star 2M0405 with stellar evolution models suggests an age of $\sim$10-25 Myr.  We measured the projected spin-orbit angle of ``b" as $\lambda=15_{-16}^{+15}$ and $\lambda = 2_{-4}^{+12}$ degrees using the apparent RV shift and change in line profile, respectively, induced by the transient occultation of the rotating star by the planet. These values indicate a prograde orbit like that of the interior ``c" planet of \hoststar\ and point to a co-planar multi-planet system that formed within a disk.  We also measured variation in the strength of the 1083\,nm triplet of neutral orthohelium as a probe of any extended/escaping atmosphere around ``b".  We detect a steady decrease in absorption over the transit which appears to arise from the star or its planetary system.  While this variation could be ascribed to ``b" or possibly to the immediately preceding transit of ``d", we cannot rule out that this is due to rapid variation in the stellar disk-integrated flux in the triplet.  The amplitude of variation ($\sim$0.04\,nm) is consistent with moderate estimates of atmospheric escape driven by XUV radiation from the central star.       
\end{abstract}




\section{Introduction}
\label{sec:intro}
Observations of planets around young stars provide ``snapshots" of planetary evolution with which we can test models of changes in planet orbits, atmospheres, and compositions.  Around the youngest stars (ages of $\lesssim10$\,Myr), many of which retain primordial circumstellar disks, the process of planet formation itself can be probed.  Pushing the planetary clock back to ages approaching 1\,Myr is seen as crucial given evidence that planet formation proceeded rapidly in the Solar System \citep[e.g.,][]{Lee1996,Kruijer2017}, and theoretical models proposing that the accretion of some types of planets can occur rapidly \citep[e.g.,][]{Ormel2017,Johansen2017}.

The youngest nearby region of ongoing star formation is the Taurus Molecular Cloud, which contains representatives of all the earliest stages of star formation, including Class 0 (deeply embedded), 0/1, and 1 (emergent protostar) sources with inferred model ages $<1$\,Myr \citep{Froebrich2005}.  Taurus is also highly structured \citep{Kraus2008,Joncour2018,Galli2019}, with multiple groups with distinct age spanning $\sim$5 Myr, and a more disperse population of older stars of poorly determined ages, but which could be at least 10~Myr old \citep{Krolikowski2021}.  Confirmation of earlier detections of giant planets around Taurus stars with the radial velocity (RV) method \citep{Donati2016,Johns-Krull2016,Yu2017} has proven elusive \citep{Damasso2020,Donati2020b}.  Direct detection by imaging has yielded companions with brown dwarf masses \citep{Itoh2005,Luhman2009}, while a candidate super Jupiter-mass planet embedded within a disk \citep{Kraus2012} remains controversial \citep{Thalmann2015,Mendigutia2018,Currie2019}.  \footnote{Gaidos et al. (in press) recently reported a Taurus star with a super-Jupiter-mass companion.}     

Space photometry missions, i.e. \ktwo\ and \tess, have opened an additional avenue for the detection of young planetary systems by the transit method.  Follow-up ground- or space-based spectroscopy during a transit can detect a planet's atmosphere in transmission, and the partial occultation of the disk of the rotating star produces a change in the shape and location of its rotationally-broadened spectral lines, constraining the angle between the stellar spin axis and orbit plane.  These phenomena may be especially pronounced for young systems where a planet's atmosphere is escaping under the influence of the active star, and the star is rapidly rotating and its lines broad and well-resolved.  The proliferation of echelle spectrographs operating in the infrared allows such observations to be made at higher signal-to-noise ratios for low-mass pre-main sequence stars like those in Taurus with \teff\ near 4000\,K.  They also provide access to the 1083\,nm triplet line of neutral orthohelium, which has been used to detect extended and escaping atmospheres \citep[e.g.,][]{Spake2018,Nortmann2018}.

\ktwo\ observed part of Taurus during Campaign 4, and \citet{David2019} reported a Jupiter-size ($10.2\pm0.6$\rearth) planet on a 24.12-day orbit transiting the solar-mass Taurus member \hoststar.  Subsequent analysis of those data revealed three additional transiting planets , two interior to and one exterior to the orbit of \planet\ \citep{David2019b}.  The radii of these planets is anomalously large compared to counterpart multi-planet systems around middle-aged ($\gtrsim$1\,Gyr) tars, suggesting that the light-element dominated envelopes of these planets are heated or still hot, and/or the atmospheres are escaping.  \citet{Feinstein2021} report observations of the ``c" planet of \hoststar\ and constrain the spin-orbit angle $\lambda$ to 5$\pm$15\,deg (a prograde orbit).  We obtained high-resolution echelle IR spectra of a transit of \planet\ to (a) measure the ``Doppler" shadow of the planet in the shape of the spectral lines and to search for anomalous absorption in the \hei\ triplet.

\section{Observations and Data Reduction}
\label{sec:observations}

An updated linear ephemeris for \planet\ (T. David, pers. comm.) predicts a transit that was partially visible from Maunakea Observatory with a mid-transit time $T_c$ at BJD = 2459119.0664 (UT 13:36 on 26 September 2020), however there are significant transit time variations (TTVs) due to mutual perturbations between the planets of \hoststar\ (Livingston et al., in prep.).  A fit to a TTV model predicts $Tc$ at BJD = 2459119.0231 or 62 minutes earlier at UT 12:33 with an uncertainty of 1.7\,min.  We monitored \hoststar\ with the the InfraRed Doppler (IRD) spectrograph \citep{2012SPIE.8446E..1TT, Kotani2018} on the 8.2-m Subaru telescope on Maunakea.  We obtained 41 spectra of \hoststar\ beginning at first contact (T1), with 3 before T1, and 41 within the 6.42 hours span between T1 and last contact (T4).  Monitoring concluded at the end of the night and 31 min before T4.  We also obtained 97 other spectra outside of transit to generate a template spectrum which is free from the telluric-line features for our RV analysis.  IRD covers 970-1730\,nm with $\lambda/\Delta \lambda \approx 70,000$. Integration times were $8-$min during and near the transit.  Using the \texttt{IRAF} echelle package \citep{Tody1986} and a custom reduction pipeline \citep{Hirano2020b}, we extracted one-dimensional spectra after flat-fielding and scattered light subtraction.  Wavelengths were calibrated using the comparison spectra of the Th-Ar lamp as well as the laser-frequency comb (LFC) taken during each run. Typical signal-to-noise (S/N) ratios at 1000\,nm were 75--115.

\section{Analysis and Results}
\label{sec:analysis}

\subsection{Systemic radial velocity}
\label{sec:rv}
The RV analysis for NIR spectra taken by IRD is described in detail in \citet{Hirano2020b}. In short, we first generated a high-SNR template spectrum of \hoststar\ that are free of telluric features, by removing telluric features and combining multiple observed IRD spectra taken in different epochs. Individual spectra were then fitted using this template spectrum, taking into account the instantaneous instrumental profile (IP) of the spectrograph inferred from the simultaneously observed LFC's spectra. RVs were extracted for each small spectral "segment"  with a wavelength range that was chosen to cover at least a few stellar absorption lines. For \hoststar, having a relatively large rotation velocity ($v\sin i>20$ km s$^{-1}$), a typical segment length of $2-3$ nm was adopted.  The typical RV internal error for each $8-$min integration was $34-45$ m s$^{-1}$. The mean systemic barycentric RV is +14.644$\pm$0.136\,\ks. 

\subsection{Taurus membership}
\label{sec:membership}

\begin{figure}
	\includegraphics[width=\columnwidth]{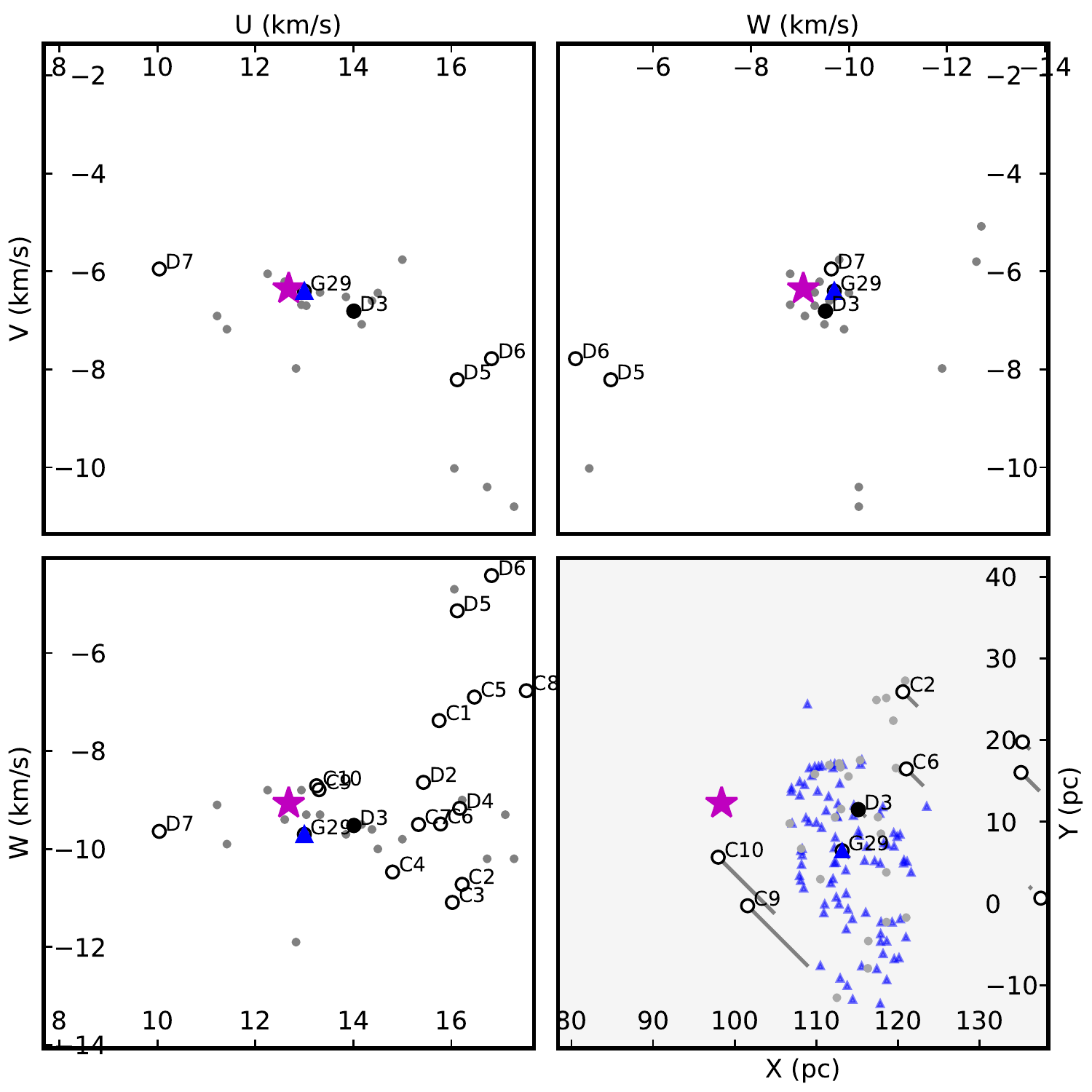}
    \caption{Upper left and right and lower left panels: $UVW$ space motion of \hoststar\ relative to some of the Taurus groups cataloged by \citet{Krolikowski2021} and Group 29 (blue triangle) identified by \citet{Luhman2018}.  We identify the dispersed D3-South sub-group (filled circle) as the most likely host group.  Lower right: Position of \hoststar\ relative to those groups in Galactic Cartesian coordinates.  Galactic $Z$ relative to \hoststar is represented as the length of the ``pins".  Individual members of the D3-South group and Group 29 are plotted as grey and blue triangles ($UVW$ are not available for the latter).  Coordinate systems are left-handed, i.e. positive $U$ and $X$ towards the Galactic center.}  
    \label{fig:uvwxyz}
\end{figure}

Using parallaxes and proper motions from the \gaia\ EDR3 data release \citep{Gaia2020} and the RV determined from IRD, we calculate $XYZ$ = ($-98.4 \pm 0.2$,$+12.26 \pm 0.03$, $-42.86 \pm 0.01$) pc and $UVW$= ($12.68 \pm 0.01$, $-6.35 \pm 0.02$, $-9.07 \pm 0.01$) \ks, using left-handed coordinates (positive $X$ and $U$ towards the Galactic center).   The star is clearly associated with the Taurus cluster, but falls immediately outside the central core of the cluster.  Comparing these to the position and kinematics of the subgroups identified by \citet{Krolikowski2021}, the dispersed \hoststar\ is kinematically indistinguishable from the ``D3-South" group among the Taurus sub-groups identified by \citet{Krolikowski2021}, but lies distinctly in the foreground of the cataloged members (Fig. \ref{fig:uvwxyz}).  \citet{Krolikowski2021} catalog the star as a member of ``D2"; this sub-group is a younger, more condensed group.  There is significant spatial overlap of D3-South with the Group 29 identified by \citet{Luhman2018}, and the centroid kinematics are very similar (Fig. \ref{fig:uvwxyz}). Estimates of their ages, however, differ markedly: 6.2$^{1.4}_{1.7}$Myr for D3-South, but with a significantly older ($\sim$13 Myr) component {Krolikowski2021}, and between the ages of the $\beta$ Pictoris and Tucana-Horologium moving groups (20-45 Myr) for Group 29 \citep{Luhman2018}.

\subsection{Multiplicity, mass and age}
\label{sec:mass-age}

We revisited the properties of the host star by fitting stellar atmosphere models to available photometry, i.e. from the \citet{Paunzen2015} catalog of Stromgren-Crawford $ubvy\beta$ photometry, APASS $BVgri$, \emph{Tycho} $B_TV_T$, \gaia\ $GB_pR_p$, 2MASS $JHK_s$, and \wise\ W1-4.  The best-fit model to 26 data points ($\chi^2=12.5$ for 18 degrees of freedom) has \teff=4900K, log g = 3.0, Fe/H = 0.  A Bayesian analysis marginalizing over all models yields average values of \teff=4911\,K and $\log g = 3.57$.  The inferred integrated flux combined with the \gaia\  parallax yields a luminosity of $0.903 \pm 0.002$\,\lsun and a radius of $1.317 \pm 0.027$\,\rsun, consistent with but more precise than \citet{David2019}.  \citet{David2019} found a rotation period of $2.865 \pm 0.012$\,days which gives an equatorial speed of $23.2 \pm 0.5$ km~sec$^{-1}$, consistent within errors with the $v \sin i$ measured from optical echelle spectra (\citealt{David2019} and \citealt{Feinstein2021}) and indicating that the stellar  rotation axis lies close to the plane of the sky.  

\begin{figure}
	\includegraphics[width=\columnwidth]{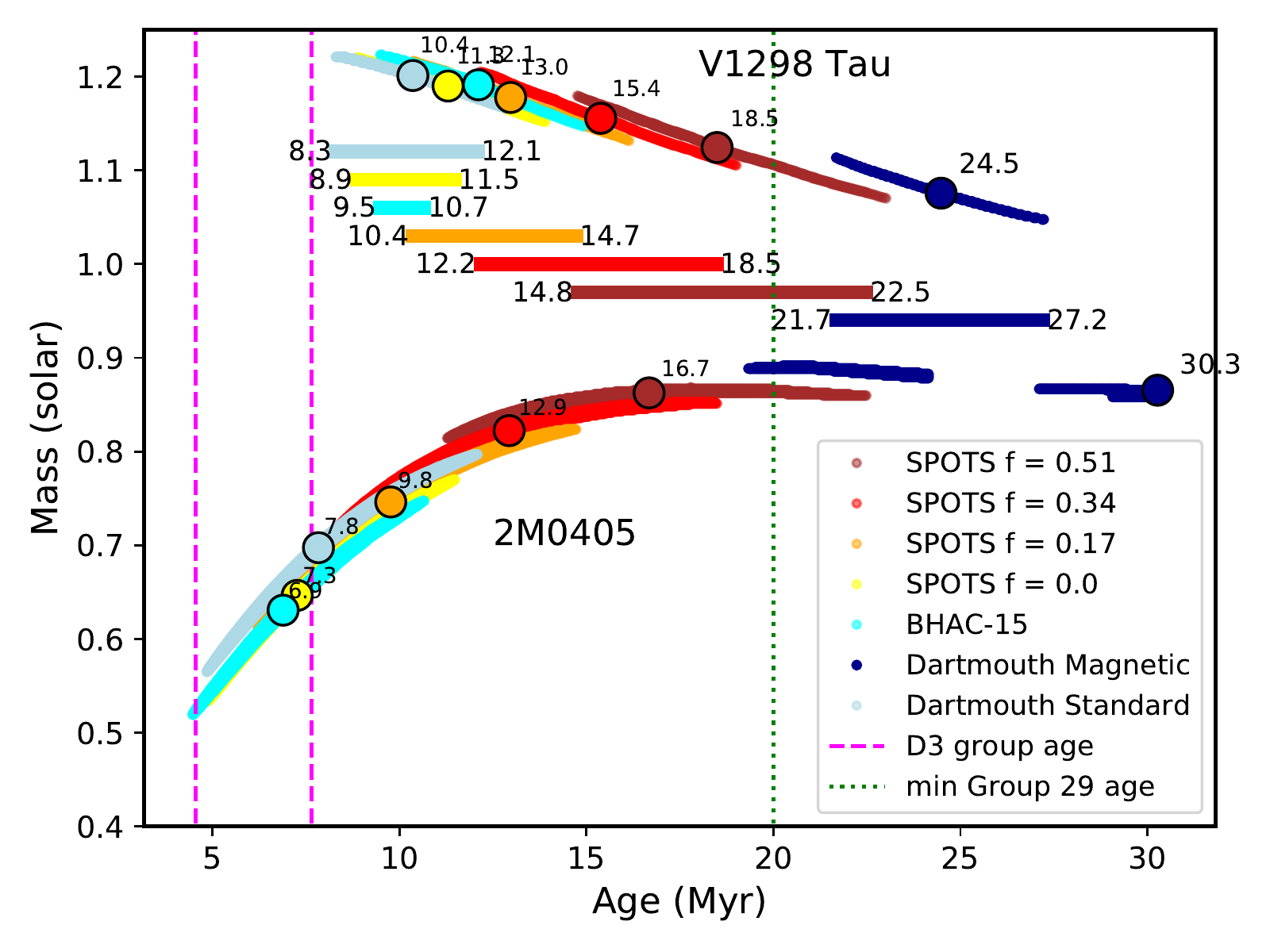}
    \caption{Masses and ages of Dartmouth \citep{Dotter2008}, BHAC-15 \citep{Baraffe2015} and SPOTS models \citep[][, for different spot coverage fractions]{Somers2020} that are consistent within a 95\% confidence level ($\chi^2 <6$ for 2 degrees of freedom) with the luminosity and \teff\ of \hoststar\ (upper loci) and its candidate co-moving companion 2M0405 (lower loci).  Each color represents those interpolated models with \teff\ and $L_{\rm bol}$ consistent with observations.  The colored circles are the best-fit (minimum $\chi^2$) values labeled with the corresponding age.  The color-coded horizontal bars represent and are labeled by the overlap in 2$\sigma$ model ages, i.e. the allowed age range if the stars are co-eval.  The dotted line is the minimum age of the co-moving Group 29 described by \citet{Luhman2018}, while the vertical dashed magenta lines are the estimated age range of the co-moving Taurus D3-South subgroup describe \citet{Krolikowski2021}.  The discontinuity in the Dartmouth Magnetic fits for 2M0405 is a result of a gap in model parameter coverage.}  
    \label{fig:mass-age}
\end{figure}

We compared the $L_*$ and \teff\ of \hoststar\ to three different sets of models: the BHAC-15 model \citep{Baraffe2015} which do not account for the effects of stellar activity; the Dartmouth models with \citep{Feiden2016} and without \citep{Dotter2008} the effect of magnetic fields that inflate low-mass on the pre-main sequence; and the SPOTS models \citep{Somers2020}, which account for the effects of spots and their magnetic fields with different coverage fraction.   (Fig. \ref{fig:mass-age}).  This comparison points to a mass of 1.07-1.22\msun, depending on model, and an age of 8-23 Myr ($\delta \chi^2$ or 95\% confidence for 2 degrees of freedom), which is intermediate between those assigned to the D3-South group and Group 29, with the heavily spotted and magnetic models yield older ages, as expected.  While the D3-South age is based on a comparison with BHAC-15 models \citep{Krolikowski2021}, the Group 29 models are based on a comparison with isochrone fits to other co-moving groups, and differences between group member selection, models and fitting procedures might underlie some of the discrepancy.  

Simultaneous fitting of multiple co-eval stars, the basis of isochrone age-dating of stellar clusters, can significantly improve mass and age estimates compared to fits of individual stars.  In the case of Taurus this is hindered by the region's young age and significant substructure and age heterogeneity.  Companion stars formed from the same molecular cloud core are more likely to be coeval since the collapse time of dense molecular cloud cores is $\sim10^5$\,yr \citep{Chevance2020}.  We therefore searched for possible companions to \hoststar\ by identifying stars with similar parallaxes and proper motions in the \gaia\ EDR3 database out to a radius of 1 deg.  Only two other stars have parallaxes within 5$\sigma$ of that of \hoststar\ and similar proper motions: HD 284154, 98\arcsec\ away, and 2MASS\,J04051277+2007489, hereafter 2M0405, with a separation of 137\arcsec.  Both stars were identified as significant X-ray emitters and possibly young and related to \hoststar\ by  \citet{Poppenhaeger2021}.  HD\,284154, was identified as a potential companion to \hoststar\ by \citet{Oh2016}, but (ironically) ruled out as a Taurus member by \citet{Gagne2018}.  It does not have a published RV, but if this were to prove to be the same as \hoststar\ then the space velocity difference could be as small as 0.3 \ks.  2M0405, which is 40\arcsec\ from HD 284154, does have a \gaia\ spectrometer measurement of RV ($12.64 \pm 0.93$ \ks) which is within error of that of \hoststar.  Under the assumption that the two stars \emph{are} at the same distance the difference in velocity is $2.0 \pm 0.89$ \ks, a 2.2$\sigma$ level of significance.  The formal probability that the velocity is less than the escape speed at the projected separation (0.38 \ks) is 0.03 \footnote{These calculations assume Gaussian, independent errors which is not the case for \gaia\ astrometry.}.  Moreover, the \gaia\ Reduced Unit Weighted Error (RUWE), a measure of the astrometric error, is 1.66 which means that the star itself is likely to be a binary \citep{Belokurov2020} and its proper motion and hence velocity errors are likely to be larger than reported in EDR3.  Obtaining echelle spectra and precision RVs and AO imaging is clearly a high priority to establish whether the system is gravitationally bound and whether 2M0405 is indeed a binary is clearly a high priority. 

We first analyzed observations of this star assuming that it is a high-contrast binary, i.e. that the companion contributes negligibly to the total source flux.  We fit stellar atmosphere models to the photometry of 2M0405 (APASS DR10, 2MASS Point Source Catalog, the ALL\wise\ catalog, and \gaia\ $GBpRp$), adopting a reddening of \ebv=0.024 \citep{David2019}.  The best-fit solar-metallicity BT-SETTL model with CIFIST solar abundances yielded $\chi^2=2.44$ for 11 degrees of freedom and \teff=3900K and $\log g$ = 4.5.  We compared stellar evolution models to the inferred \teff\ and luminosity of 2M040, yielding an age of 5-22.5\,Myr.  Next, we computed concordant ages as the overlap between the individual 95\% confidence intervals.  These span much narrower ranges, with spot-free model ages spanning 8-10~Myr and progressively more magnetic models ranging up to 22-27 Myr (horizontal bars in Fig. \ref{fig:mass-age}).  The spot fraction $f_c$ and magnetic field of \hoststar\ are not known and could vary with stellar cycles.  \citet{Morris2020} model the \ktwo\ Campaign 4 lightcurve of \hoststar\ to estimate $f_S \approx 9^{+1}_{-2}$\%, which would suggest a concordance age of $\approx$10-12\,Myr (assuming a similar $f_S$ for 2M0405), significantly younger than the $23 \pm 4$\,Myr estimated by \citet{David2019}..  But it has long been known that such estimates suffer from model degeneracies \citep[e.g.,][]{Luger2021} and higher spot fractions and thus older ages are possible.  Furthermore, since 2M0405 is a binary in which each component is less luminous than the sum, its model ages could be older.  Concordance between the age estimates for the two stars is maximized by artificially reducing the luminosity of 2M0405 by 20\%, and becomes 10-15\,Myr.  Better constrains on the age of \hoststar\ will require precise characterization, binary identification, and isochrone fitting of 2M0405 and HD 284154, any other co-moving neighbors, and members of the D3-South sub-group and Group 29.     

\subsection{Rossiter-McLaughlin effect and Doppler-shadow analysis}
\label{sec:rm}

\begin{figure}
	\includegraphics[width=\columnwidth]{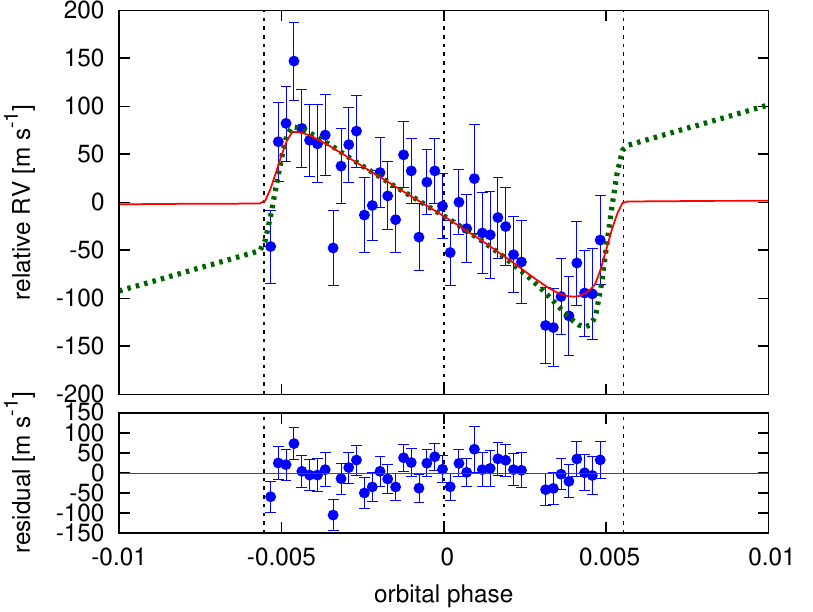}
    \caption{Relative RV of \hoststar\ on UT 26 September 2020. The best-fit RM plus RV-baseline model, in which we imposed priors on $K$ and $v\sin i$, is drawn by the red solid line. The green dotted line indicates the best-fit model, in which no priors are assumed for $K$ and $v\sin i$ (those parameters show very broad posterior distributions; see the text).
    The lower panel plots the RV residual from the best-fit model. The three vertical dashed lines in each panel represent the transit ingress ($T_1$), center ($T_c$), and egress ($T_4$) times of \hoststar\ b. }  
    \label{fig:RV}
\end{figure}

In order to constrain the stellar obliquity of \hoststar\ relative to the normal of planet b's orbit plane, we analyzed its RV sequence during the transit observed by IRD.  The relative RVs during the transit are plotted in Fig. \ref{fig:RV} and suggests an anomaly typical of the Rossiter-McLaughlin (RM) effect for transiting planets \citep[e.g.,][]{2005ApJ...631.1215W}. However, a fit to the observed RVs, in which all the relevant parameters were allowed to vary freely, resulted in poor constraints on the fitting parameters.  Due to the lack of out-of-transit baseline data on the transit night, the MCMC analysis performed poorly, leading to degeneracy among the system parameters including the RV semi-amplitude $K$, which accounts for any change in RV during the transit but unrelated to the transit itself, $v\sin i$, and the projected stellar obliquity $\lambda$ (see the green dotted line in Fig. \ref{fig:RV}, representing the best-fit RV model without any priors on the fitting parameters). Moreover, \hoststar\ is known to exhibit large activity-induced RV noise (``jitter") on the timescale of stellar rotation \citep{David2019b}, and therefore RV data taken on other nights were not helpful enough to determine the RV baseline during the transit. 
We thus opted to impose Gaussian priors on $v\sin i$, the scaled semi-major axis $a/R_s$, full transit duration $T_{14}$, and $K$ as $v\sin i = 23 \pm 2$ km s$^{-1}$, $a/R_s = 27.0\pm 1.1$, $T_{14}=6.42\pm 0.13$ hours \citep{David2019b, David2019}, and $K = 0 \pm 400$ m s$^{-1}$, respectively. \citet{2019RNAAS...3...89B} obtained $K = 10 \pm 48$ m s$^{-1}$ for \planet\, but since it was based on a relatively short-term RV monitoring, we conservatively imposed a weak prior whose width is twice as large as the size of the reported RV jitter \citep[$200$ m s$^{-1}$,][]{David2019b}.

We modeled the RV anomaly during the transit using the analytic RM formula in \citet{Hirano2011}, and implemented the MCMC analysis to estimate the obliquity $\lambda$. In addition to the above fitting parameters, we allowed the RV zero point $\gamma$ for our RV data set and the mid-transit time $T_c$ to float, for the latter of which we imposed Gaussian prior based on the predicted ephemeris (i.e., $T_c = 2459119.0231 \pm 0.0012 $ day in BJD). Implementing the MCMC simulations, we obtained $K=(-0.3\pm3.2)\times 10^2$ m s$^{-1}$, $v\sin i=23.4\pm 1.9$ km s$^{-1}$, and $\lambda=15_{-16}^{+15}$ degrees, implying a low projected obliquity for \hoststar. The best-fit RV curve with the analytic RM formula is drawn by the red solid line in Fig. \ref{fig:RV}. 

\begin{figure*}
	\includegraphics[width=15.5cm]{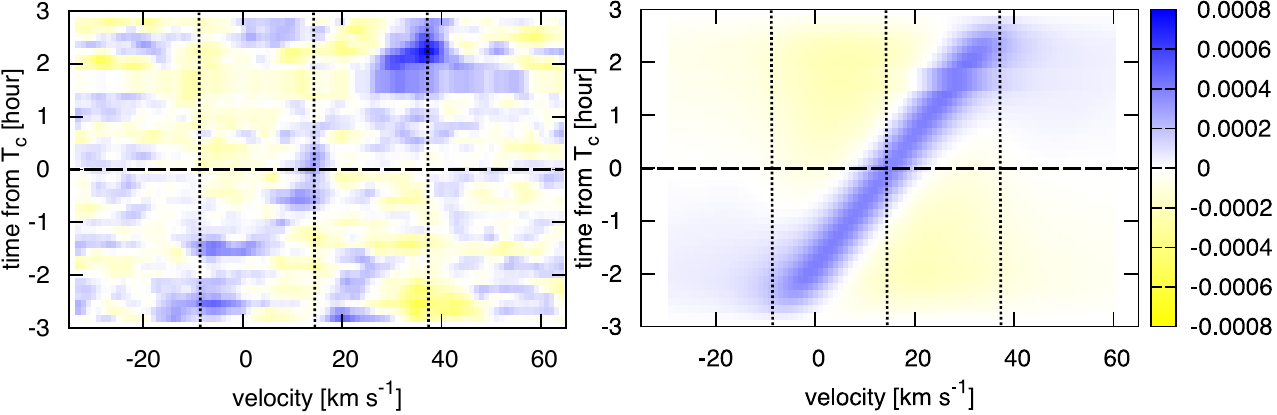}
    \caption{Residual CCF maps for the observed IRD spectra (left) and best-fit theoretical model (right). The three vertical lines for each panel indicate the approximate CCF center and its $\pm 23$ km s$^{-1}$ ($\approx v\sin i$), corresponding to the line edges. The horizontal dashed line represents the mid-transit time $T_c$. }  
    \label{fig:DT}
\end{figure*}

To further investigate the alignment between the rotation of \hoststar\ with the orbit of ``b", we investigated the variation in the line profile during the transit, performing the "Doppler-shadow" analysis as in \citet{Hirano2020, Gaidos2020b}. This technique directly detects and models the profile-variation of the cross-correlation function (CCF) of the observed spectra, caused by the partial occultation of the rotationally broadened stellar line profile. Following \citet{Hirano2020}, we cross-correlated each observed IRD spectrum of \hoststar\ against a theoretical template of \hoststar\ \citep{2005A&A...443..735C}, after correcting for telluric features. The resulting CCFs were aligned to the same reference RV taking into account the barycentric corrections of the Earth's motion. Usually, the mean "out-of-transit" CCF is subtracted from the individual CCFs to highlight the residual CCF variation, but since we lack such out-of-transit frames taken on the same night, we median-combined all the CCFs that were taken during the transit. We then subtracted this combined CCF from each CCF finally to obtain the residual CCF map as shown in the left panel of Fig. \ref{fig:DT}. While the residual CCF map includes a correlated-noise pattern, likely caused by temporally evolving stellar surface inhomogeneity, the residual CCF map indicates a relatively clear shadow of the transiting planet that is traversing the stellar disk from the red-shifted edge to the blue-shifted edge. The symmetric pattern of the shadow is indicative of the prograde, aligned orbit of \hoststar\ b. 

We estimated the projected obliquity $\lambda$ by fitting this residual CCF map. The detail of the analysis is given in \citet{Hirano2020}. In brief, we generated a number of theoretical CCFs by making "mock" IRD spectra during \hoststar\ b's transit, and the observed residual CCF was fitted using those theoretical models with $\lambda$ and other relevant parameters being optimized. Following the exact procedure as for the observed data, we combined the theoretical CCF models during the transit and subtracted the combined model from individual CCF models when comparing the observed residual CCF map with the theoretical one. This analysis yielded $\lambda = 2_{-4}^{+12}$ degrees, which is fully consistent with and more constraining than the RV analysis. The best-fit theoretical residual CCF model is shown in the right panel of Fig. \ref{fig:DT}. 

\subsection{1083~nm Orthohelium triplet}
\label{sec:hei}

The \hei\ triplet at $\approx$1083\,nm appears in a single order (1075.67-1089.84\,nm in air) of IRD spectra.  Spectra obtained during the transit of ``b" and those obtained for an independent RV monitoring program (C. Beichman, pers. comm.) were evaluated based on overall signal and interference of telluric emission (OH) and absorption lines (H$_2$O) with the \hei\ triplet.  (IRD spectra are obtained through a single fiber on the target and this precludes simultaneously measurement of the sky) We retained 53 spectra spanning 10.2 days, including all spectra obtained during the transit, from 1.1 days pre-$T_c$ to 9.1 days post-$T_c$, with adequate signal and with an observer Doppler shift $>0$ \kms\ such that telluric interference was not a concern.  Spectra were shifted to the rest-frame of \hoststar.  Potential signals from the planet in spectra obtained during transit were corrected for the planet's orbital motion (assumed circular) by dividing these spectra by a mean out-of-transit spectrum, shifting the residual spectra by $2\pi R_* t/(P \tau) = 2.74(t/\tau)$\,\kms, where $t$ is the time relative to $T_c$ and $\tau$ is the transit duration, and adding the Doppler-corrected residuals back to the out-of-transit reference.  

\begin{figure}
	\includegraphics[width=\columnwidth]{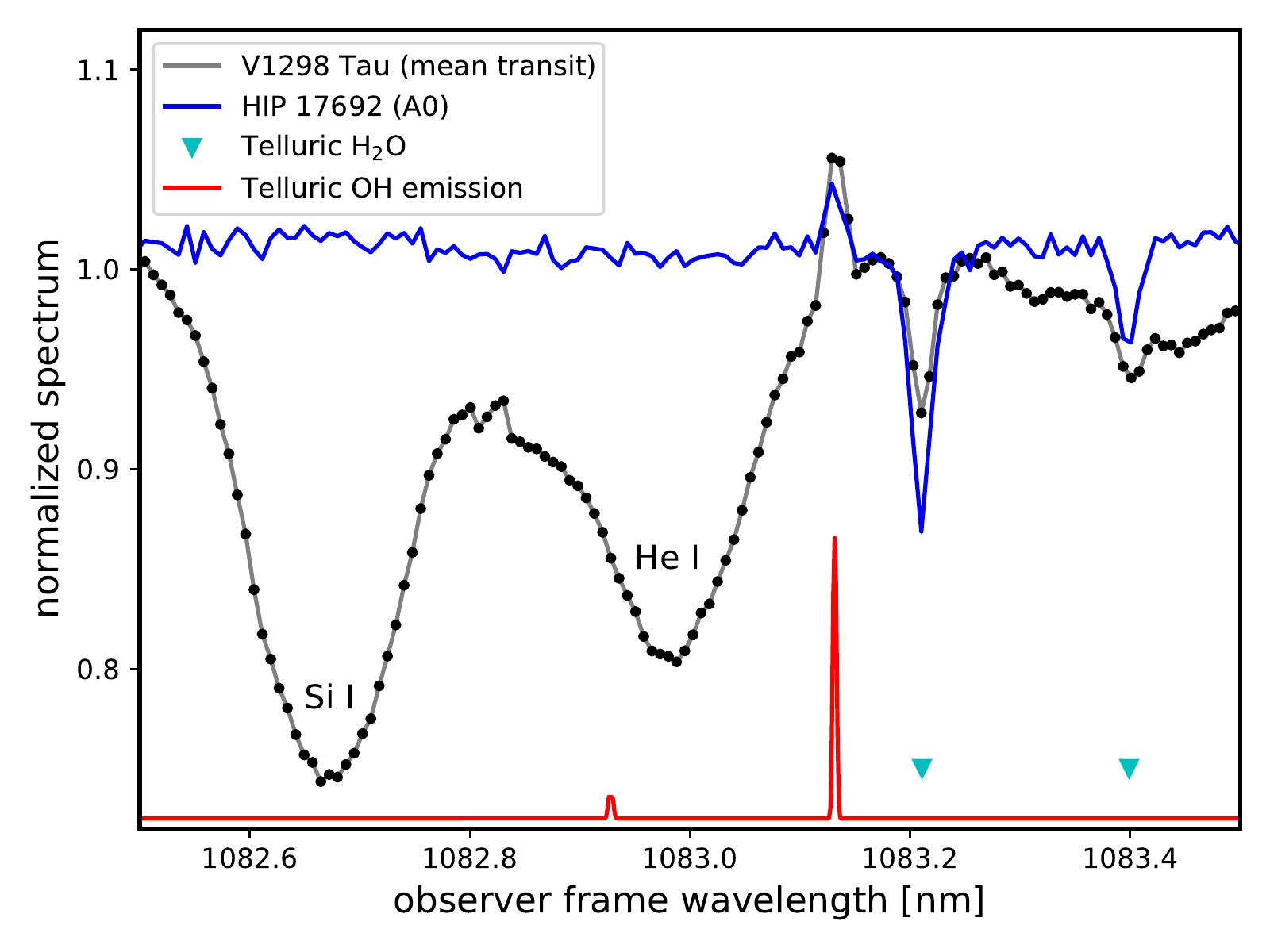}
    \caption{Average of spectra in the vicinity of the \hei\ 1083\,nm triplet of \hoststar\ obtained during the transit of ``b" with wavelengths \emph{in the frame of the observer}.  The relatively featureless spectrum of an A0-type star HIP\,17692 is also shown.  Blue triangles are telluric H$_2$O lines cataloged by \citet{Breckinridge1973} and the red curve is a model of sky emission (i.e. OH) based on \citet{Noll2012} and \citet{Jones2013}.}  
    \label{fig:hei_observed}
\end{figure}

Figure \ref{fig:hei_observed} plots the mean observed spectrum of \hoststar\ obtained during the transit, as well as the spectrum of the A0 star HIP\,17692, along with the positions of known telluric lines due to H$_2$O (in absorption) and OH (in emission).  The rapid rotation of \hoststar\ partially blends the \hei\ triplet with a nearby Si\,I line.  The two H$_2$O lines as well as the stronger, redder OH line are readily apparent in the spectra while the weaker OH line is not.  Importantly, there is no evidence for additional telluric lines.

\begin{figure}
	\includegraphics[width=\columnwidth]{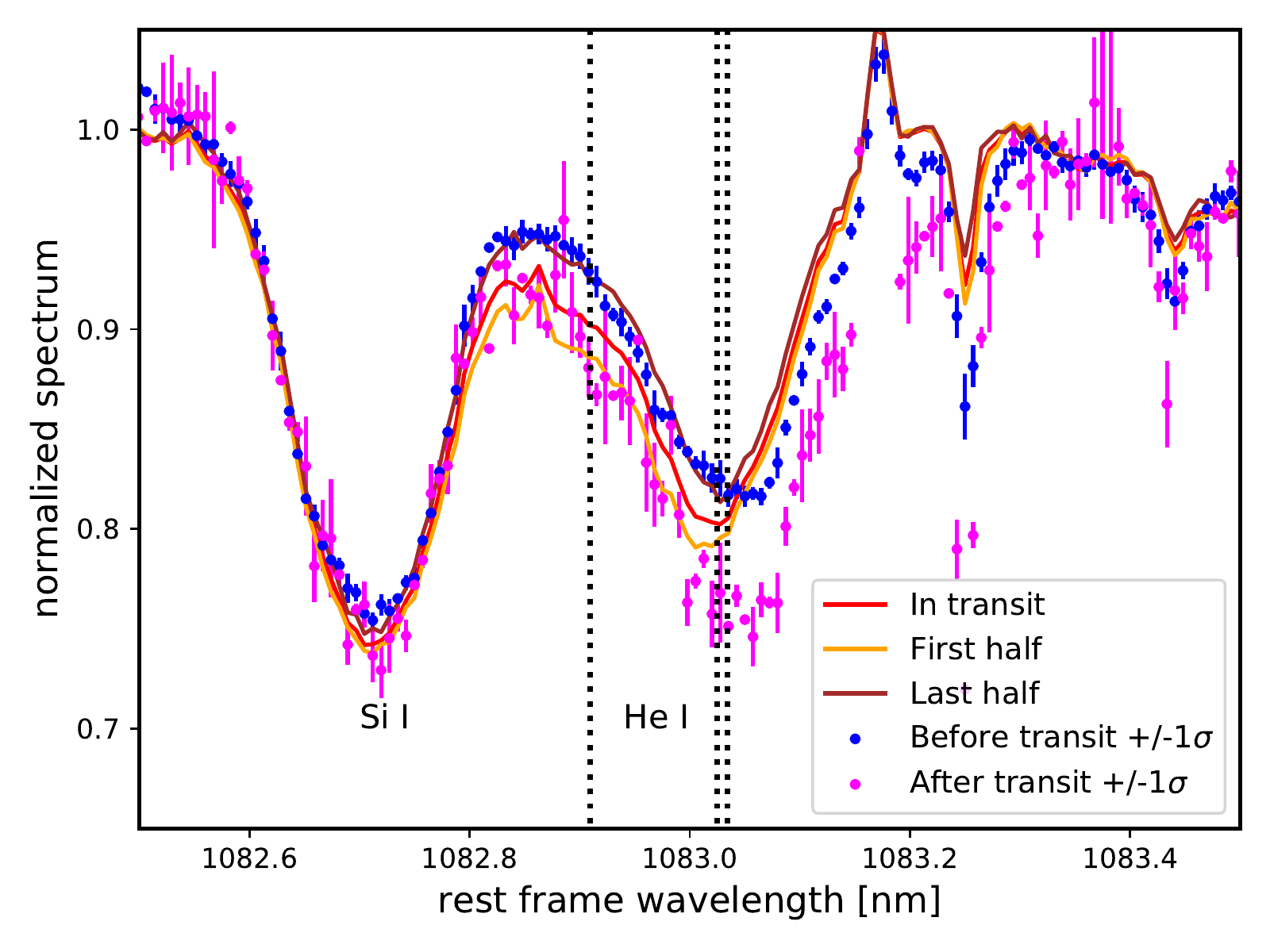}
    \caption{Spectra of \hoststar\ in the frame of the \emph{star}, showing variation in the \hei\ line before, during, and after the transit of ``b" (but no significant variation in the neighboring Si\,I line.  Error-bars show the standard deviation among the multiple spectra obtained on nights previous to and following the transit night.  Also plotted are the means of spectra obtained during the first half and last half of the transit.}  
    \label{fig:hei_restframe}
\end{figure}

\begin{figure}
	\includegraphics[width=\columnwidth]{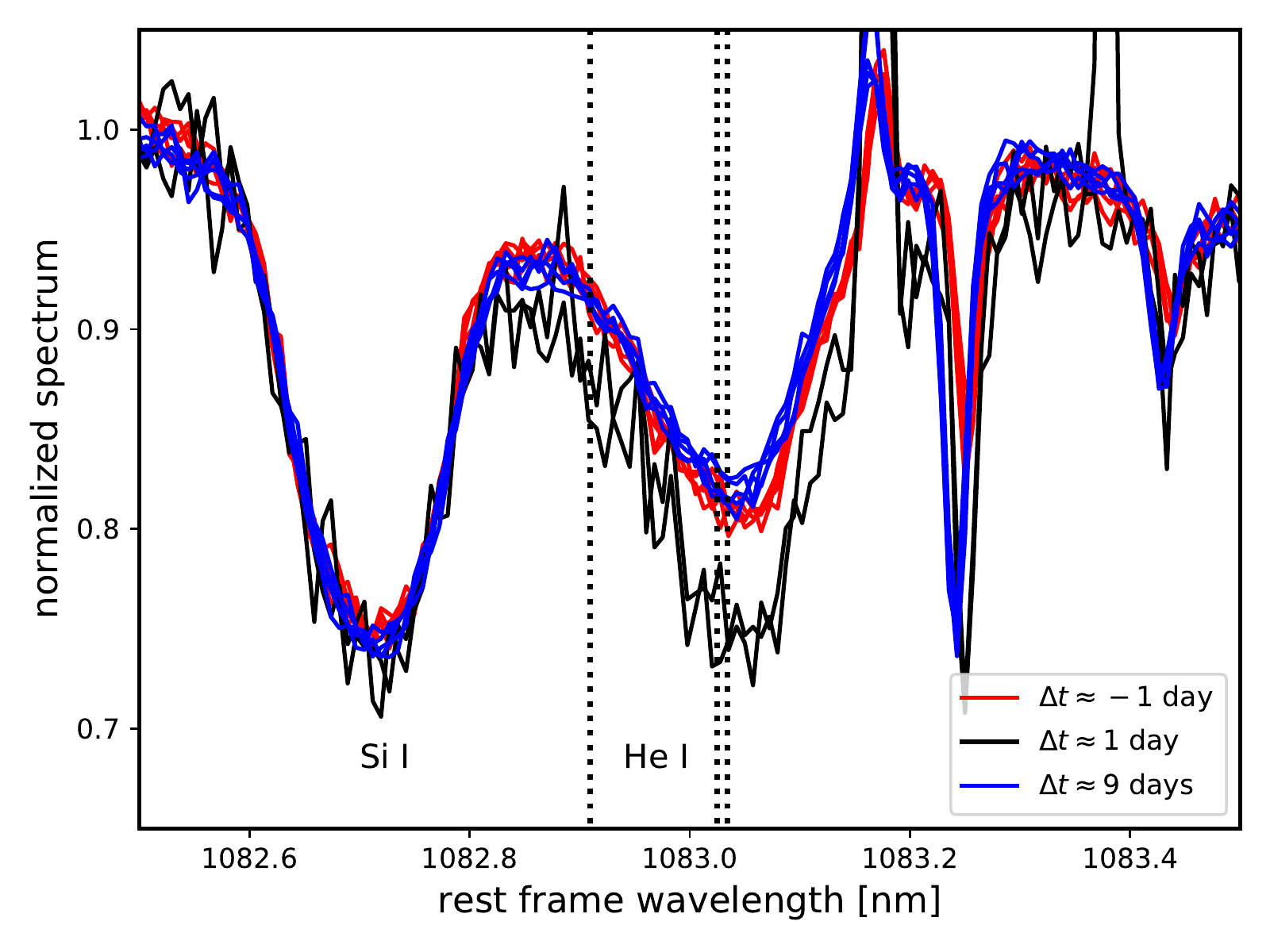}
    \caption{All spectra with acceptable signal level obtained outside of the transit of ``b" where the observed RV is $>$0 and the telluric OH line does not interfere with the \hei\ line.  The spectra are color-coded according to the approximate time relative to $T_c$.}  
    \label{fig:hei_var}
\end{figure}

Figure \ref{fig:hei_restframe} shows \hei\ spectra \emph{in the rest-frame of the star}, and compares the average of spectra obtained during the transit night, and the preceding.  The spectra during transit are also binned according to whether they were obtained during the first or second half.  There is (i) elevated absorption during the transit relative to one day before the transit; (ii) a trend of decreasing absorption and return to pre-transit absorption during the transit; and (iii) anomalously high absorption in the $N=2$ spectra obtained on the night following the transit.  In contrast, the neighboring Si\,I line is unchanged throughout this interval.  All the spectra obtained outside of transit are plotted individually in Fig. \ref{fig:hei_var}.  All are consistent except for the two spectra obtained on the following night.        

\begin{figure}
	\includegraphics[width=\columnwidth]{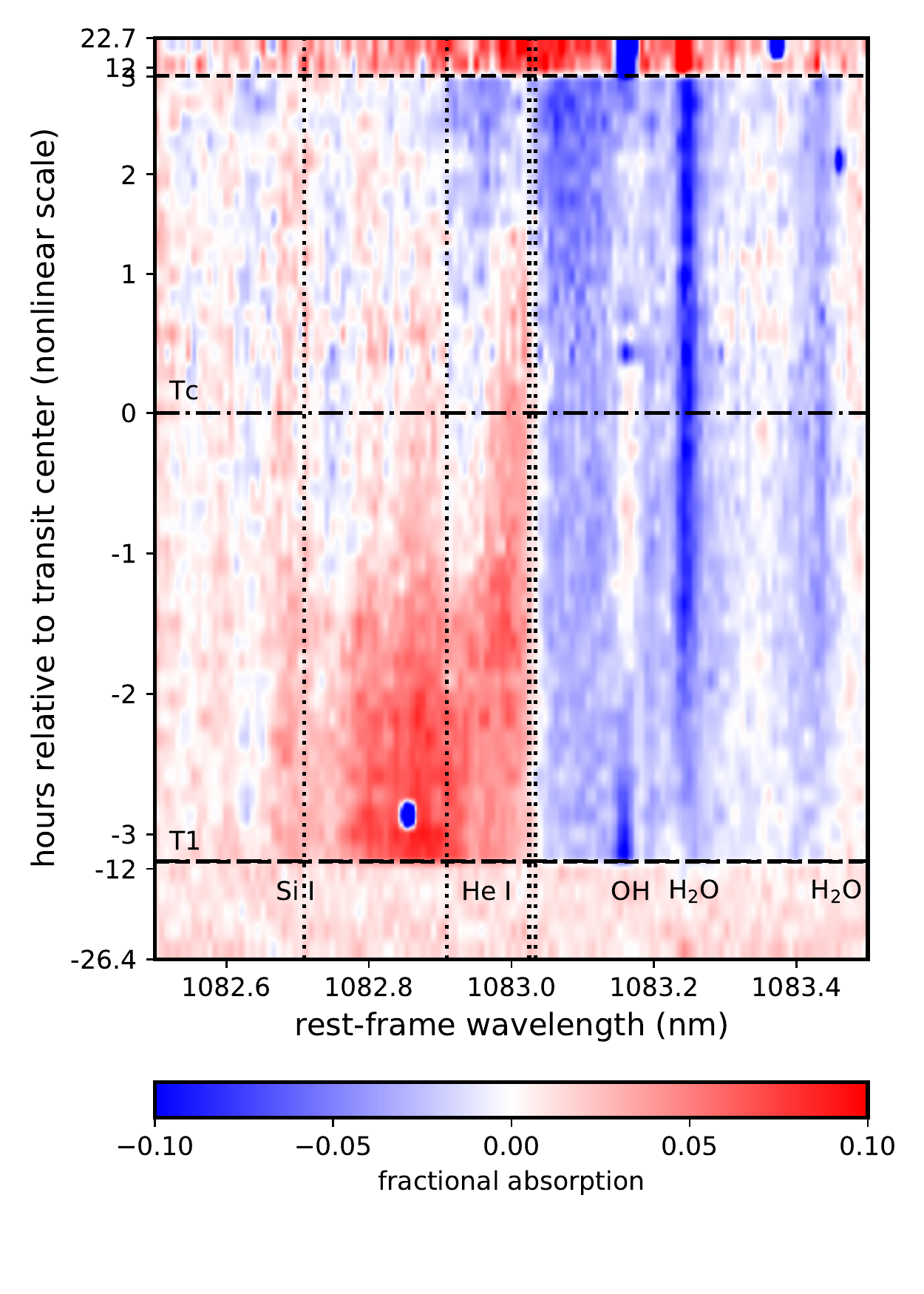}
    \caption{``Tomography" of the differential normalized spectrum of \hoststar\ in the vicinity of the \hei\ triplet relative to the before-transit mean.  Red indicates less signal (greater absorption) and blue indicates more signal (less absorption or greater emission). Each observation is represented by a separate row of pixels and since monitoring was not continuous during this period the vertical time-axis is also not uniform or continuous.  Horizontal dashed lines separate observations on different nights, while the dot-dashed line marks the beginning of the transit T1 and transit center $T_c$ (observations ceased prior to the end of transit).  Vertical dotted lines mark the individual \hei\ line centers and a neighboring \sii\ line.  Telluric lines of OH and H$_2$O are also marked.  The Doppler shift of the planet's rest-frame is negligible on this scale.  }  
    \label{fig:tomography}
\end{figure}

Figure \ref{fig:tomography} is a ``tomographic" plot of the deviations of the \hei\ line during the transit of ``b" relative to the pre-transit profile (mean of observations on the previous night).  Red areas are ones of decreased signal (increased absorption) and blue areas represent increased signal (decreased absorption).  This plot clearly shows the increase in the strength of both the singlet and doublet which make up the \hei\ triplet which begins sometime between observations of the previous night and the start of observations around transit first contact (T1), the decline in the absorption over the night, the anomalous absorption on the following night, and the return to pre-transit level 8 nights later.  There also appears to be a slight increase in emission at the longer wavelength side of the triplet.  Variability in the telluric lines of OH and H$_2$O, well separated from the \hei\ are evident, as well as the lack of variability in the \sii\ line.

We measured the total equivalent width (EW) of the He\,I triplet, the neighboring \sii\ and \mgi\ lines (1082.7089 and 1081.1076 nm), and the Paschen $\beta$ line of H\,I at 1281.8072 nm.  Lines were modeled with Voigt profiles, line constants were obtained from the NIST Atomic Line Database, and the FWHM of the Gaussian component of the Voigt profile was fit as a free parameter.  Two additional free parameters in the \hei\ fit were the ratio of the singlet to the doublet line strength (to allow for variation due to regions on the star's surface where the lines are optically thick and thus the ratio can change) and an overall Doppler shift relative to the NIST line center (to account for an uneven distribution of of active regions -- a significant source of \hei\ absorption -- on the rotating surface of the star.

\begin{figure}
	\includegraphics[width=\columnwidth]{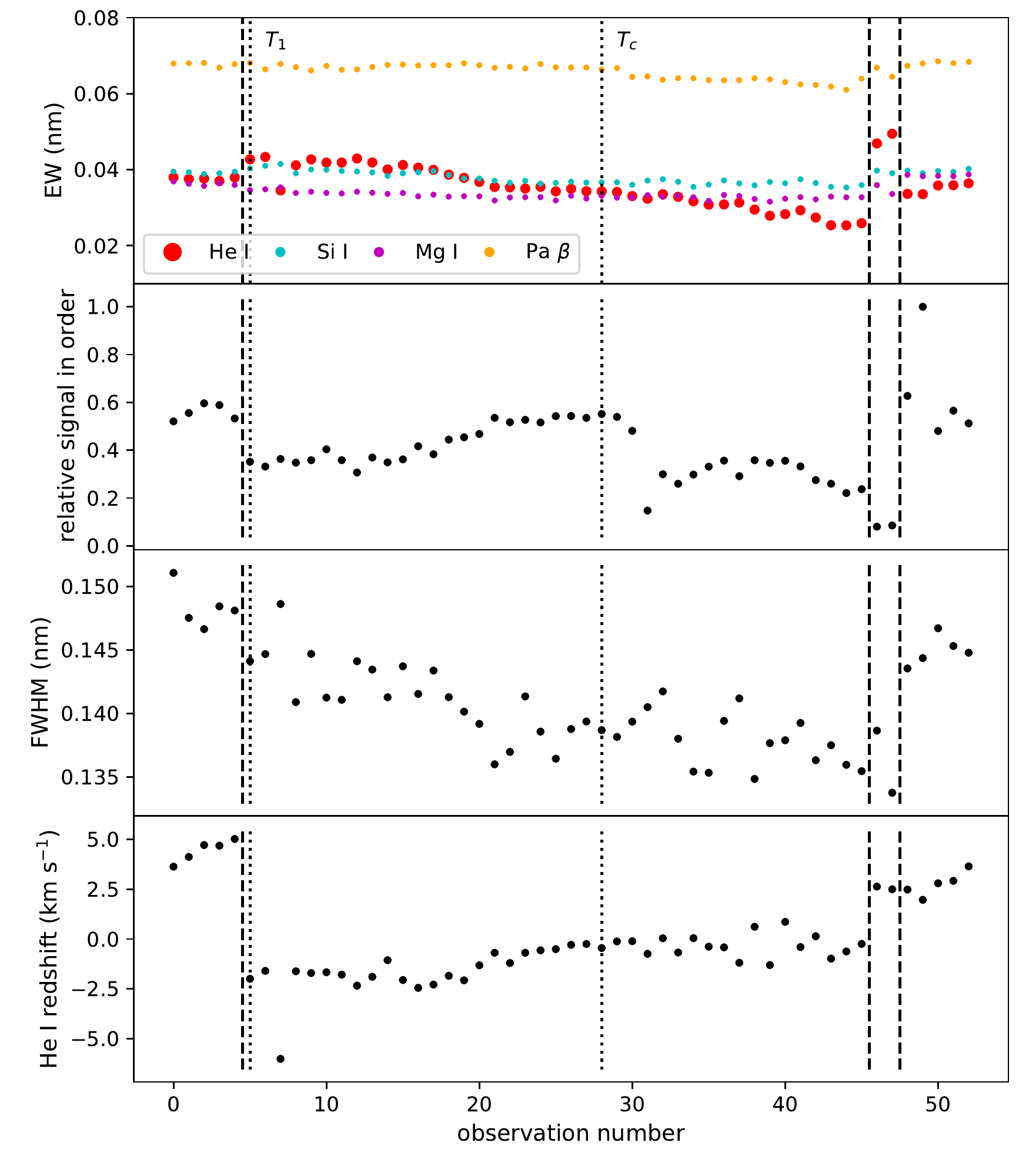}
    \caption{Top row: EW of the \hei\ triplet, neighboring Si\,I and Mg\,I lines, and the Paschen $\beta$ line of H\,I (a potential indication in stellar activity).   All data with sufficient signal and non-interference with tellurics are plotted.  Dashed lines separate different nights and the beginning ($T_1$) and mid-point ($T_c$) of the transit of ``b" are marked by dotted lines.    Second row: relative signal in the spectral order containing the \hei\ triplet.  Third row: FWHM of the He I triple and Mg I and Si I lines as a fit parameter in the Gaussian component of a Voigt line profile model.  Bottom row: red-shift in the \hei\ triplet relative to the NIST values as a fitting term in the line profile model.}  
    \label{fig:ew}
\end{figure}

The top panel of Figure \ref{fig:ew} shows the EW vs. time obtained during the night of the transit, the preceding and following nights, and 9 nights after the transit.  While the \hei\ EW measured one day before and 9 days after the transit are consistent, the the EW varies by $\sim$50\% during the transit, decreasing from 0.047 to 0.028 nm.  The EW of other lines in the neighborhood of 1083.2 nm vary by much less.  This variation is not correlated with the overall signal in the spectral order (second panel of Fig. \ref{fig:ew}), although the high \hei\ EW values on the night after the transit were obtained with lower signal during cloudy conditions.   The FWHM of the \hei\ line (third panel of Fig. \ref{fig:ew}) exhibits a slight decrease during the transit and returns to the pre-transit value, indicating that the transit-associated absorption is broader than the out-of-transit line.   During the transit, the best-fit line center (bottom panel of Fig. \ref{fig:ew}) is blue-shifted by $\sim$4 \ks\ during the transit relative to outside the transit; this is clearly seen in the blue side of the \hei\ line profile in Fig. \ref{fig:hei_observed}.

Possibly, the observed changes in \hei\ EW could have been due to variation in the intensity of the known, weaker OH line (Fig. \ref{fig:hei_observed}) or an unknown H$_2$O line.  However, the observed variation appears much broader than that observed for other OH/H$_2$O lines and nothing analogous appears in the relevant wavelength range of either \hoststar\ or the A0-type telluric reference star HIP\,17692.  Error in determination of the continuum level would manifest itself in the inferred EW, and is a leading contender to explain the drift in the \mgi\ and \sii\ lines, but this would effect all neighboring lines equally and thus cannot explain the behavior of the \hei\ line.  Variation in the instrument profile is minimal based on the FWHM behavior as well as the small (10-12 m~sec$^{-1}$) instrumental drift (highly correlated with FWHM) during the transit and to first order is unlikely to explain this variation.

Astrophysical explanations for the variation during the transit include changes in the line as a result of stellar activity, i.e. a flare and  rotation of active regions into and out of the visible disk of the star.  \hoststar\ is a rapidly rotating (2.87-day), active star and these effects are expected to be enhanced relative to middle-aged stars.  Strong \hei\ line emission and/absorption can also arise from inner disk winds or accreting disk gas \citep{Edwards2003,Fischer2008,Kwan2011} but \hoststar\ lacks the infrared excess of a T Tauri-like disk or any sign of accretion \citep{David2019}.

\hei\ emission from a flaring region fills in the line core and produces a weakening both of \hei\ at 1083 nm and the D3 line at 587.6\,nm that tracks other chromospheric emission indicators such as H$\alpha$ \citep{Schmidt2012,Johnson2021}.  However, the change in the steady decline in EW over the 6~hr observation interval is inconsistent with the \emph{increase} expected in the declining phase of a flare, and the time scale is inconsistent with the more rapid ($\lesssim$1\,hr) rise time of flares.\footnote{Flare-induced dimming, perhaps a result of electron-collisional ionization of triplet \hei\ has been observed on the Sun \citep{Kerr2021} but is low contrast and localized and unlikely to be responsible.}  Emission is also associated with Doppler shifts of 10s of \ks\ \citep{Johnson2021} which we do not observe. 

The rotation period of \hoststar\ has been determined to be 2.87 days and thus over the 6 hour observing window approximately 30$^\circ$ of longitude appears and disappears, i.e. 1/6th of the visible disk.  A 50\% variation would require a fortuitous arrangement of active regions on the stellar surface. Notably, values well before and after transit agree within a few \% (upper left panel of Fig. \ref{fig:ew}). \citet{Feinstein2021} found no variation in the \hei\ D3 line during a 6-hour observation of a transit of the ``c" planet, but did detect a steady 6\% decrease in H$\alpha$ absorption which they attributed to rotation or variation of active regions on the star, and well within the range observed among other young stars (their Fig. 7). \citet{Fuhrmeister2020} found no evidence for rotational variability of the 1083\,nm \hei\ line among M dwarfs (including a few rapid rotators), although this might be due in part to limitations of their sample, and of course \hoststar\ is young and much more active than the typical middle-aged M dwarf.  While we cannot rule out a stellar cause, other observations cast doubt on this as the explanation for a 50\% variation in \hei.

\begin{figure}
	\includegraphics[width=\columnwidth]{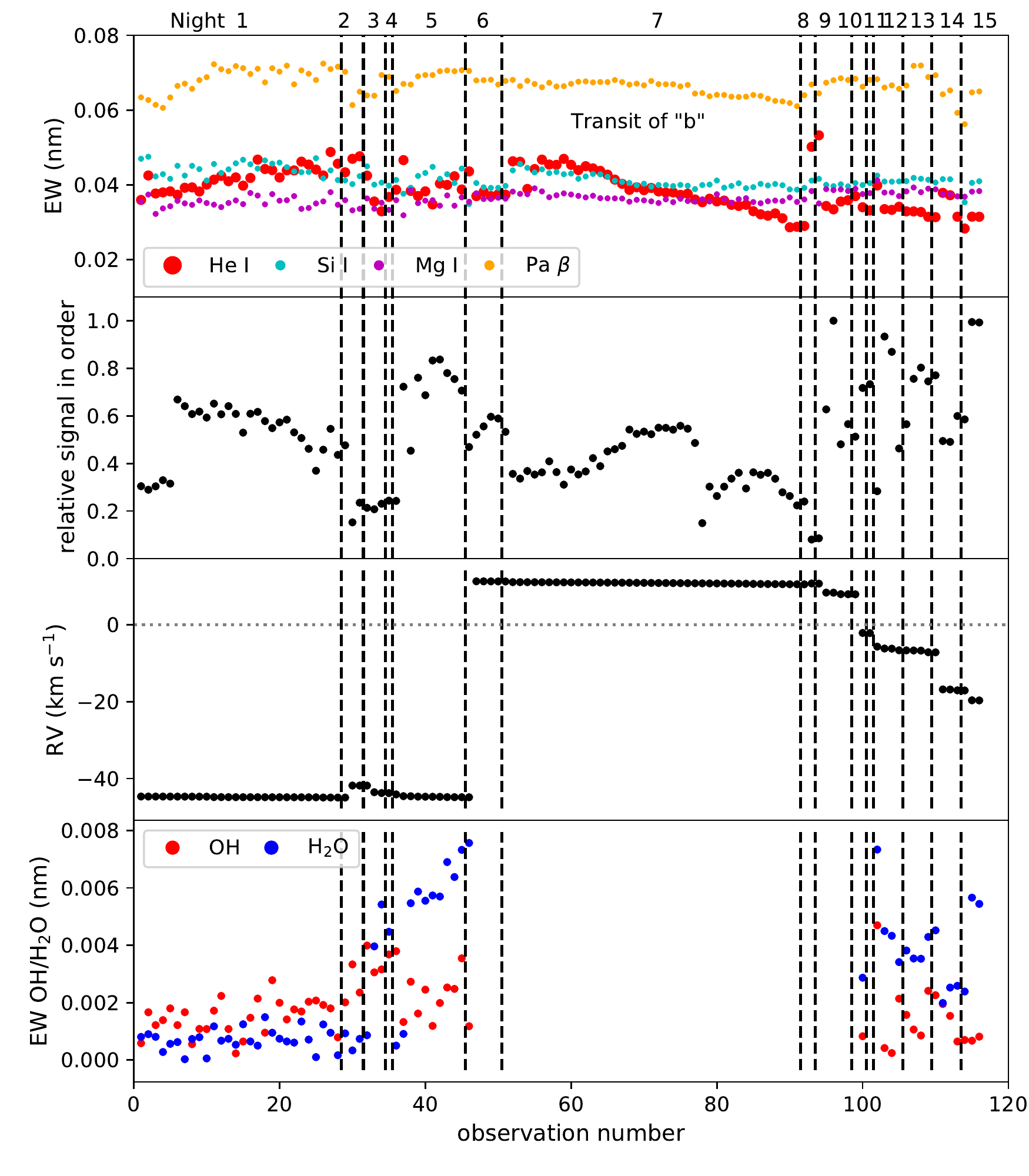}
    \caption{Similar to Fig. \ref{fig:ew} but for all spectra with acceptable signal, along with diagnostics of spectrum quality.  Data are plotted in simple temporal order, with vertical dashed lines separating different nights, which are numbered at the top.  Top row: EW of the \hei\ triplet, neighboring Si\,I and Mg\,I lines, and the Paschen $\beta$ line of H\,I (an indicator of stellar activity) determined by fitting Voigt line profiles.  Second row: relative level of signal in the order containing the \hei\ triplet.  Third row: Observer-frame RV of the star, which determines interference by tellurics.  In spectra obtained with RV $<0$ (below the dotted grey line) the \hei\ triplet is contaminated by telluric OH and/or H$_2$O lines.  Bottom row: EW of the OH (red, emission) and H$_2$O (blue, absorption) lines estimated by fitting Gaussian line profiles simultaneously with stellar lines.  In spectra obtained with RV $>$0, the OH and H$_2$O are outside the fitting region and EWs are not plotted.}  
    \label{fig:ew_all}
\end{figure}

We further investigated the intrinsic variability of the stellar \hei\ line by analyzing all spectra with sufficient signal, regardless of telluric contamination, by simultaneously fitting the OH and H$_2$O lines with Gaussian profiles with a fixed FWHM of 0.0155\,nm (the instrument resolution) but variable equivalent width and line center, along with the \hei\ triplet and blended \sii\ line.  While these fits capture the overall strength of the lines, the limited spectral sampling and departures of the line profiles from idealized models degrade the fidelity of the fits.  The top panel of Fig. \ref{fig:ew_all} shows that there is apparent variability of a magnitude similar to that observed during the transit, but this occurs when the observer-frame RV of the star (third panel of Fig. \ref{fig:ew_all}) means that the \hei\ line is contaminated by telluric OH and H$_2$O lines, and the \hei\ EW variation is largely correlated with the strength of those lines (bottom row of Fig. \ref{fig:ew_all}).  In particular, the apparent drop in \hei\ EW on nights 3-5 and 10-15 relative to nights 1-2 occurs when the interfering telluric lines are significantly stronger.  The increasing trend over 3 hours on night 1 suggests that the star could be responsible for the variation we seen during the transit on night 7 \footnote{No known planets transited during night 1 and the next transit (of "c") started about 16 hours after these observations ended.}. 

The alternative astrophysical explanation is an affect associated with the ``b" planet, but this is difficult to conclusively demonstrate with these data.  The occultation of individual active regions by the transiting planet will produce variation, but since the transit depth is 0.6\% \citep{David2019}, these regions would have to occupy only $\sim$0.3\% of the stellar disk, which is completely inconsistent with the observed strength of the \hei\ line.  Occultation by an extended He-containing cloud would seem to be required, but the cloud would have to \emph{precede} the planet in its orbit.  Moreover, the anomalous absorption on the night following the transit would remain to be explained.  The formal uncertainty in $T_c$ is only $\pm$2 min.

\hoststar\ hosts multiple planets: could another of these be responsible?  Using the linear ephemerides for the \hoststar\ system (Livingston et al., in prep), a transit of ``c" occurs with a $T_c$ at BJD = 2459118.3044 (UT 19:18 on 25 September), 17.25 hr before that of "b", and a transit of ``d" has a $T_c$ at BJD = 2459118.6915 (UT 4:36 on 26 September), only 7.96 hours before ($\pm$8 min).  \emph{The transit of ``d" ends less than two hours before the beginning of our observations for the transit of ``b".}  This raises the possibility that the declining \hei\ absorption observed during the transit of ``b" is actually associated with the prior transit of ``d".   The radius of ``d", $6.4 \pm 0.4$\rearth, is only marginally larger than that of ``c" ($5.6 \pm 0.3$\rearth), and both almost certainly have low-molecular weight (H/He-rich) envelopes.  Why the more irradiated ``c" has little or no mass loss \citep{Feinstein2021} but ``d" might is unclear, but one possibility is simply that the yet-to-measured mass and thus gravity of ``d" is much lower than that of ``c".

\section{Summary and Discussion}
\label{sec:discussion}

The study of very young planetary systems such as that of \hoststar\ offer insight into the timescales and mechanisms of planet formation and evolution that shape the exoplanet populations that large surveys such as \kepler, \tess, and RV surveys have revealed.  One outstanding question is whether planets on close-in orbits formed close to their host star, migrated inward via torques from the natal disk, or were scattered inwards by gravitational interaction with another planet or companion star.   The last scenario predicts that the normal of the planetary orbit plane could be significantly inclined to the stellar rotation axis, particularly at early times before tidal torques from the star could re-align the orbits.  The observations of \citet{Feinstein2021} and our work presented here that the projected obliquity of the ``c" and ``b" planets, respectively, is small (and possibly zero) point to an origin involving the protoplanetary disk.

The low stellar obliquity for the \hoststar\ system is in line with the earlier results of obliquity measurements for young systems \citep[e.g.,][]{2020ApJ...892L..21Z, 2020AJ....160..193D, Hirano2020, Gaidos2020b}. This is also consistent with the empirical finding that cool exoplanet hosts ($T_\mathrm{eff}\lesssim 6250$ K) generally have low stellar obliquities \citep{2010ApJ...718L.145W, 2012ApJ...757...18A}; this empirical law was found for giant planets ($\gtrsim 5\,R_\oplus$) like \planet. The number of obliquity measurements for young exoplanet systems is still limited ($<10$), and further measurements for young exoplanets covering a wide range of stellar types would enhance our understanding on the origin and dynamical history of close-in exoplanets. 

\citet{Poppenhaeger2021} use models of XUV-driven H/He escape to predict that the innermost (``c" and ``d") could experience significant or near-complete loss of their H/He envelopes in 5 Gyr. Current mass-loss rates depend sensitively on the heavy-element (``core") content, but \citet{Poppenhaeger2021} estimate it could be as high as 0.8 \mearth\ in 10 Myr.  An important caveat is that they adopt the energy-limited formulation with a constant efficiency factor of 10\%, which \citet{Krenn2021} use detailed models of hydrodynamic escape to show can lead to errors of orders of magnitude in escape rates.

The EW of the \hei\ triplet is elevated during a transit of ``b" relative to nights before and after the transit, and shows a steady decreasing trend during the transit.  This does not seem to be a product of the instrument, telluric effects, or our analysis.  It is difficult to understand this in terms of a surrounding and/or trailing cloud of neutral gas around ``b", but our observations begin only 1 hour after the transit of ``d", a smaller (and presumably less massive) planet on an interior orbit.  Multiplying the $\approx$2 hr duration of the absorption event plus the 1 hr between the end of the transit of ``d" and the beginning of our observations, with the orbital velocity of ``d" yields a transverse extent of $\approx 7 \times 10^5$ km, which is about 26 planet radii or 1.1 stellar radii.  Due to its size, such a cloud would have to be optical thin in the \hei\ line.    

\citet{Vissapragada2021}, also report no detection of \hei\ absorption in a narrow \hei-band filter associated with transits of ``b" but do find tentative evidence for absorption associated with partials transits of ``d" on UT 8 Oct and 27 Nov 2020.  The corresponding EW of their measurement is $0.027 \pm 0.014$ nm, which is consistent with our measurement of an $\approx$0.19 nm change during the 26 September transit.  They also obtained spectra containing asymmetric variation of the \hei\ line on day-timescales that could be driven by stellar flaring.

\begin{figure*}
	\includegraphics[width=\columnwidth]{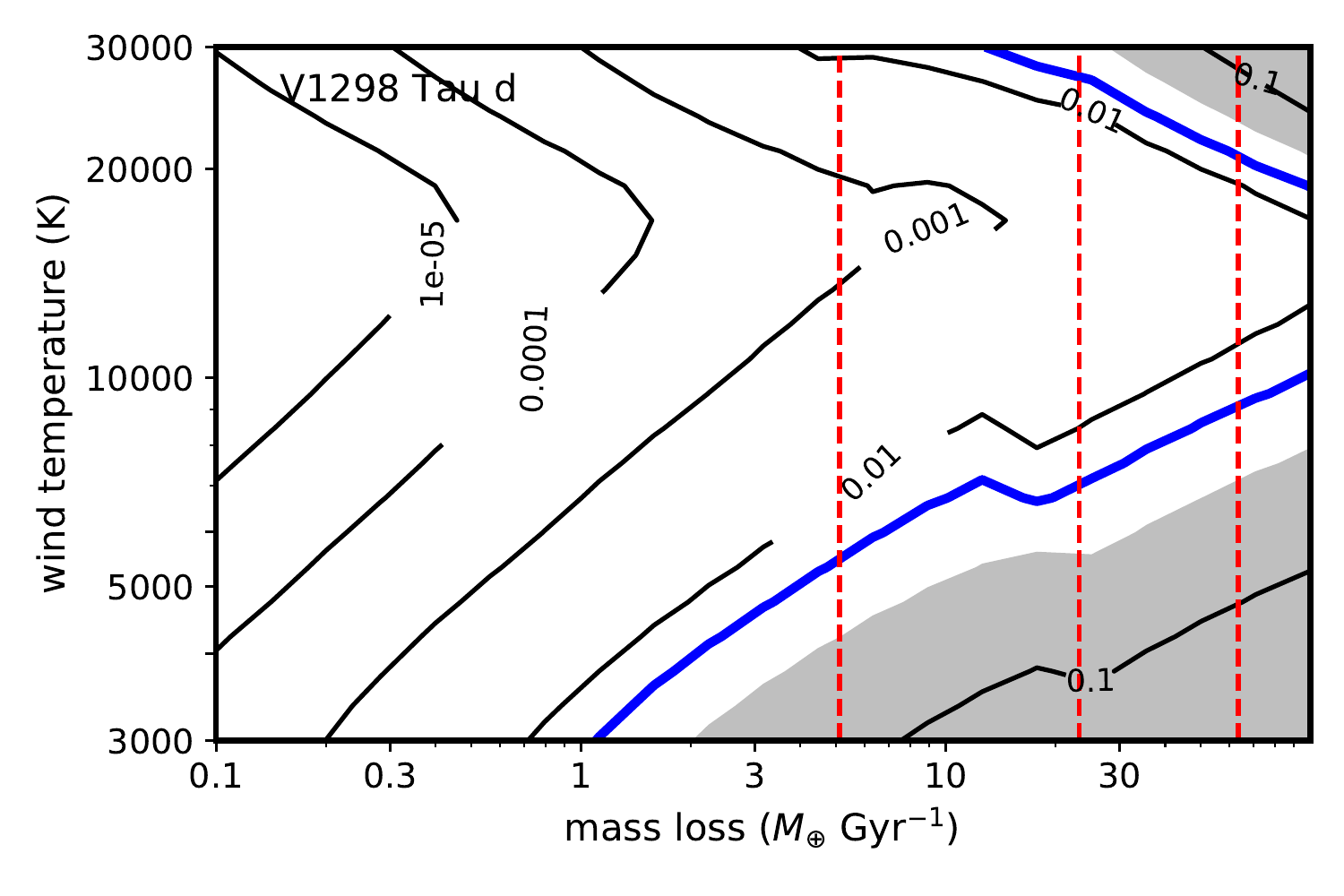}
	\includegraphics[width=\columnwidth]{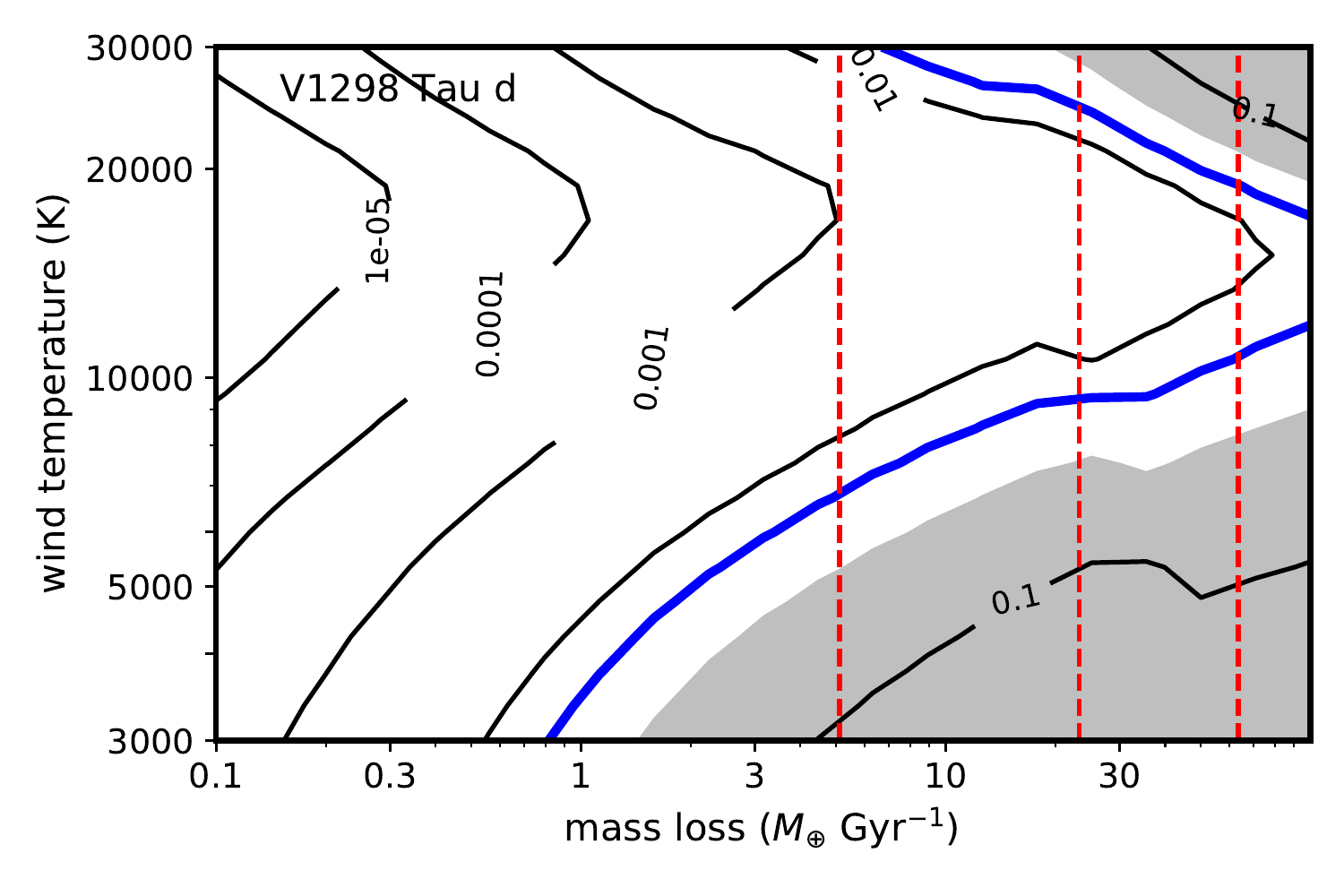} 
    \caption{Contours of constant \hei\ (in nm) predicted by a model of atmospheric escape as an isothermal Parker wind from \hoststar\ b (left) and d (right).  The thick blue contour represents the 0.019 nm change observed during the transit of ``b".  The grey region where EW $>$0.04\,nm can be excluded by our observations.  The dashed red lines mark the low, medium, and high loss-rate estimates of \citet{Poppenhaeger2021}.}  
    \label{fig:model}
\end{figure*}

We estimate the expected \hei\ absorption produced by an escaping atmosphere of H/He from ``b" or ``d".  The model described in detail elsewhere \citep{Hirano2020,Gaidos2020a,Gaidos2020b}, calculates the escaping atmosphere as an isothermal (temperature $T_{\rm wind}$ Parker wind with a solar-like composition (H/He=10.5).  The spectral intensity from the star in the extreme ultraviolet (EUV) is a required input, since photons with $\lambda < 50.4$\,nm are responsible for ionizing \hei\ and, via recombination, producing the neutral triplet state.  Also needed is the intensity in the near ultraviolet (NUV), since photons with $\lambda < 258.3$\,nm ionize the triplet state.  The UV spectra of distant stars in molecular clouds are not possible to measure directly due to absorption and scattering by intervening gas and dust and must be inferred from other measures such as X-rays.  We adopted the spectrum of the young, active K2-type star Epsilon Eridani as reconstructed by the MUSCLES Treasury Survey \citep[v. 2.2,][]{France2016,Youngblood2016,Loyd2016}.  The spectrum was adjusted by the square of the radius ratio of \hoststar\ to Eps Eri.  The EUV portion of the spectrum was also further adjusted so that the 0.01-1 kEV EUV ($\lambda\lambda$ 12.39-123.9\,nm) luminosity is $5 \times 10^{30}$ ergs sec$^{-1}$, as estimated by \citet{Poppenhaeger2021}.  The latter was done by choosing a value for \lyalpha\ emission such that adjusting individual EUV wavelength ranges using the empirical relations between \lyalpha\ and EUV emission of \citet{Linsky2014} produces the \citet{Poppenhaeger2021} estimate for XUV luminosity.  For a given planet radius, the predicted mass loss rate depend sensitively on planet mass, but for a given escape rate the \hei\ signal also depends only weakly on planet mass through the atmospheric scale height; for both ``b" and ``d" we assumed 10\mearth.  Figure \ref{fig:model} shows the predicted equivalent width in nm as a function of mass loss rate and wind temperature, along with the observed change of 0.019 nm during the transit (thick blue line) and excluded levels $>0.04$ nm (grey zone).  Assuming that the observed change in \hei\ EW is \emph{not} associated with ``b", our observations rule out some or all of the range of mass loss estimated by \citet{Poppenhaeger2021}, if $T_{\rm wind}<$7000K (Fig. \ref{fig:model}a).  If instead the change is produced by an extended wind from ``d", the inferred escape rate (Fig. \ref{fig:model}b) would support the predictions of \citet{Poppenhaeger2021} and result in the loss of an entire $\lesssim$1\mearth\ H/He envelope in $\sim$100 Myr.   

Our findings highlight the promise but also the challenges of observations of planets orbiting active young stars.  In the case of \hoststar, our results and those of \citet{Feinstein2021} place limits on escape of the atmospheres of the b and c planets, respectively; extending this to the ``d" planet is a high-priority next step.


\section*{Acknowledgements}

We thank Trevor David for providing an early version of the transit ephemerides, and Greg Feiden for Dartmouth magnetic stellar model tracks.  EG was supported by NASA Grant 80NSSC20K0957 (Exoplanets Research Program).   This work is partly supported by JSPS KAKENHI Grant Numbers JP20K14518, JP19K14783, JP21H00035, JP18H05442, JP15H02063, and JP22000005.  We used NASA's Astrophysics Data System Bibliographic Services, the Centre de Donn\'{e}es astronomiques de Strasbourg, NIST's atomic line database, {\tt Astropy} \citep{Astropy2013}, and {\tt Scipy} \citep{Scipy2019}.\\

{\bf Data Availability:}  All data used in this work are available from the authors or the Subaru SMOKA archive.





\bsp	
\label{lastpage}
\end{document}